# The Bridge Toward 6G:
# 5G-Advanced Evolution in 3GPP Release 19

Xingqin Lin

NVIDIA

Email: xingqinl@nvidia.com

*Abstract*— **The 3rd generation partnership project (3GPP) initiated 5G-Advanced in Release 18, laying a solid foundation for the further evolution of 5G-Advanced. Release 19–the next wave of 5G-Advanced–will primarily focus on commercial deployment needs while serving as a bridge toward 6G. In this article, we provide an in-depth overview of the 5G-Advanced evolution in 3GPP Release 19. We not only delve into the key technology components and their corresponding use cases in 5G-Advanced evolution but also shed light on initial 3GPP studies toward 6G.**

## I. INTRODUCTION

Since its debut in Release 15 developed by the 3rd generation partnership project (3GPP), the fifth generation (5G) of mobile communication has focused on three primary usage scenarios: enhanced mobile broadband (eMBB), ultra-reliable low-latency communications (URLLC), and massive machine type communications (mMTC) [1]. 5G supports both non-standalone (NSA) operation, utilizing long-term evolution (LTE) for initial access and mobility, and standalone (SA) operation without reliance on LTE. Subsequently, 3GPP has been dedicated to advancing 5G technology evolution through Releases 16 and 17 to enhance performance and accommodate new applications [2].

Release 18 signifies the commencement of work on 5G-Advanced, which is built upon the foundation laid by 3GPP in Releases 15, 16, and 17 for 5G [3]. Existing features, such as multiple-input multiple-output (MIMO), multicast and broadcast service (MBS), dynamic spectrum sharing (DSS), quality-of-experience (QoE), non-terrestrial network (NTN), carrier aggregation (CA), integrated access and backhaul (IAB), and reduced capability (RedCap), have been further enhanced. New features addressed in Release 18 include artificial intelligence (AI)/machine learning (ML) for next-generation radio access network (NG-RAN), network energy savings, operation with less than 5 MHz bandwidth, uncrewed aerial vehicle (UAV), and extended reality (XR), among others. Release 18 also investigates new areas for future evolution, such as AI/ML for new radio (NR) air interface, duplex operation, and low-power wake-up signal (LP-WUS) and low-power wake-up receiver (LP-WUR) [4]. Figure 1 outlines 3GPP's roadmap for the evolution of 5G from its inception to 5G-Advanced.

The substantial enhancements and explorations conducted in Release 18 lay a solid foundation for the evolution of 5G-Advanced in Release 19. The discussion on the scope of Release 19 started at the 3GPP RAN Release-19 workshop, which attracted about 480 submissions from over 80 companies/organizations [5]. It was endorsed at the workshop that 5G-Advanced in Release 19 would primarily focus on commercial deployment needs while serving as a bridge toward the sixth generation (6G) of mobile communication. After six months of intense deliberation, 3GPP endorsed the Release-19 package at the December 2023 RAN plenary meeting [6]. This package comprises diverse but selective study and work items aimed at enhancing network performance, addressing new use cases, and acting as a stepping stone toward 6G.

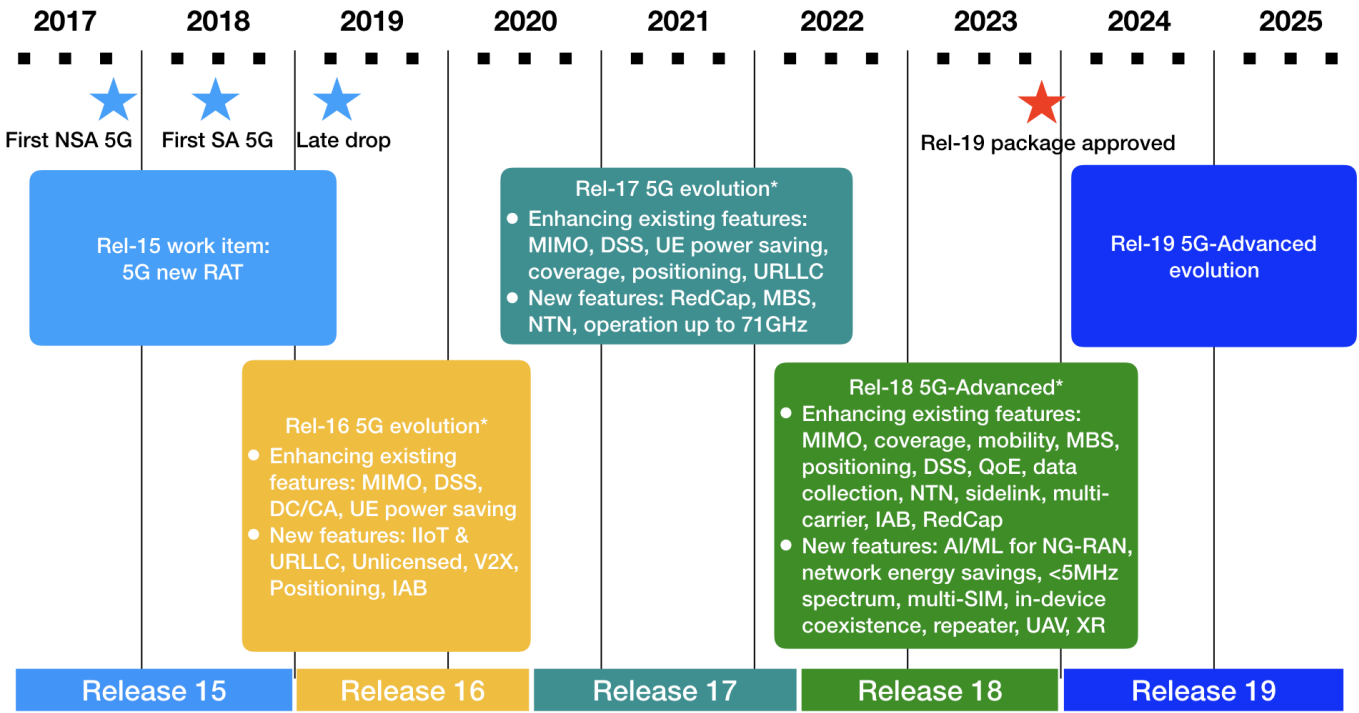

Figure 1: 3GPP's 5G evolution roadmap from 5G to 5G-Advanced (indicative).

This article provides a holistic overview of the 5G-Advanced evolution in 3GPP Release 19. It delves into the primary technology components and their corresponding use cases, while also shedding light on initial 3GPP studies toward 6G. In particular, we categorize the 17 approved/endorsed items into five groups in this article and present them in the following five sections titled high performing 5G, further evolved 5G topology, energy efficient 5G, AI-powered 5G, and toward 6G, respectively.

## II. HIGH PERFORMING 5G

3GPP will further enhance 5G performance in Release 19 by introducing new functionality in the areas of MIMO evolution, mobility management, duplex operation, and XR support, as illustrated in Figure 2.

MIMO is an intrinsic, pivotal technology component in 5G [7]. It has been continuously enhanced over 5G releases and will enter phase 5 evolution in 3GPP Release 19. In the current beam management framework, a 5G node B (gNB) can request a user equipment (UE) to carry out beam reporting to facilitate the gNB in determining a suitable beam for communicating with the UE. It is however difficult to balance between reporting overhead and latency/accuracy. To mitigate this issue, 3GPP will specify UE-initiated beam reporting enhancements to allow the UE to trigger beam reporting rather than waiting for the gNB to trigger. 3GPP will also increase the number of ports for channel state information (CSI) reporting from 32 to 128 to better support larger antenna arrays which have attracted increased interest in the industry. Another enhancement area is coherent joint transmission (CJT) in the scenarios with non-ideal synchronization and backhaul such as inter-site CJT, for which UE measurement and reporting of inter transmit-receive point (TRP) time misalignment and frequency/phase offset will be introduced. With the emergence of UEs equipped with three transmit antennas, 3GPP will enhance non-coherent uplink codebook to facilitate codebook-based transmissions with three antenna ports. Last but not least, enhancements will be introduced to better support heterogeneous networks, where a UE may receive transmission from a macro gNB in the downlink but transmit to one or more micro TRPs in the uplink to increase uplink throughput. To support such asymmetric downlink single TRP and uplink multi-TRP scenarios, two closed-loop power control adjustment states for sounding reference signals (SRS), where one is used for the macro gNB and the other is used for micro TRPs, will be needed. Besides, it will be necessary to configure pathloss offset for the UE to calculate uplink pathloss to micro TRPs based on downlink reference signal from the macro gNB.

Mobility support is another pivotal technology component for high-performing 5G networks. In Release 18, 3GPP introduced a new mobility procedure known as layer 1 (L1)/layer 2 (L2)-triggered mobility (LTM), which can reduce handover latency and interruption time compared to the classical mobility procedure based on layer 3 (L3) measurements and radio resource control (RRC) signaling. However, the Release-18 LTM procedure can only be used for mobility between cells associated with the same central unit (CU) of a gNB, i.e., intra-CU LTM. In Release 19, 3GPP will extend the intra-CU LTM procedure to support mobility between cells associated with different CUs, i.e., inter-CU LTM. 3GPP will also introduce event triggered L1 measurement reporting for LTM to reduce signaling overhead compared to periodic L1 measurement reporting. Besides, the L1 measurements for LTM will be extended from synchronization signal block (SSB) measurements in Release 18 to include CSI reference signal (CSI-RS) measurements in Release 19 to improve the LTM performance. Another major area for Releas-19 mobility work is to introduce the support of conditional LTM. The basic idea of conditional handover is that the network sends a handover command to the UE with a condition and the UE applies the handover command only when the condition is satisfied. Conditional handover can improve mobility robustness as it reduces signaling exchange during handover. Conditional LTM will enjoy both reduced interruption time brought by LTM and improved mobility robustness brought by conditional handover.

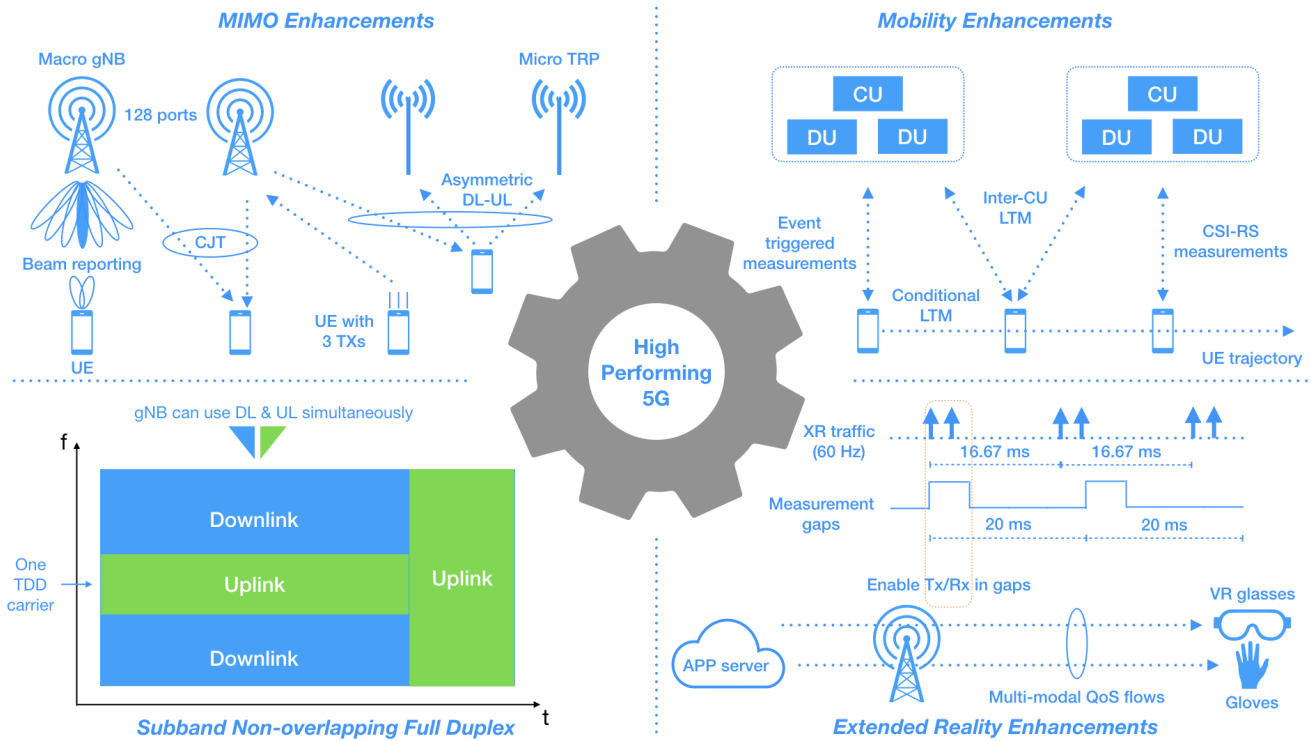

**Figure 2: An illustration of the key 3GPP Release-19 features related to high performing 5G.**

Time division duplex (TDD), where the entire frequency spectrum can be used for either downlink or uplink at different time instances, is widely deployed in commercial 5G networks. In Release 18, 3GPP carried out a study on subband non-overlapping full duplex (SBFD) at the gNB side within a conventional TDD carrier [8]. SBFD allows simultaneous co-existence of downlink and uplink at the same time within the TDD carrier, leading to a mix of TDD and frequency division duplex (FDD). In particular, configuring an uplink subband within the time duration of legacy downlink symbols in the TDD carrier can extend the uplink time duration, resulting in improved coverage, increased capacity, and reduced latency. Following the conclusion of the Release-18 study on SBFD, 3GPP will specify SBFD operation at the gNB side within a TDD carrier in Release 19. The work will entail indication of time and frequency locations of SBFD subbands to UEs, where an SBFD subband refers to consecutive resource blocks that share the same transmission direction. The work will also include transmission, reception, and measurement behaviors and procedures associated with the SBFD operation, as well as support of random access in SBFD symbols. Another enhancement area is the handling of cross-link interference (CLI) which occurs when a node is transmitting while another node is receiving in the same frequency band. CLI exists in dynamic/flexible TDD, and the SBFD operation will compound the CLI problem. Both gNB-to-gNB and UE-to-UE co-channel CLI handling schemes will be considered in Release 19.

XR use cases demand high-performing 5G networks that can simultaneously provide high data rates as well low and bounded end-to-end latency. In Release 18, 3GPP introduced several enhancements for XR support in 5G networks, such as XR awareness in the RAN [9]. In Release 19, 3GPP will continue enhancing 5G performance to support XR use cases by enabling transmission/reception in gaps or restrictions that are caused by radio resource management (RRM) measurements and improving radio link control acknowledged mode operation. Besides, enhancements for packet data convergence protocol as well as uplink scheduling with delay information will be studied and specified if justified. 3GPP will also investigate techniques to facilitate efficient and effective support for XR use cases with multi-modal quality-of-service (QoS) requirements.

## III. FURTHER EVOLVED 5G TOPOLOGY

5G needs to be deployed in diverse deployment scenarios in order to meet the requirements of different consumer and enterprise services, calling for flexible network topology. 3GPP Release 19 will continue to study and evolve 5G topology by exploiting diverse technologies including 5G femtocell, wireless access backhaul (WAB), multi-hop sidelink relay, and NTNs, as illustrated in Figure 3.

A femtocell, deployed at home or at enterprise premises, is a wireless access point used to enhance indoor cellular connectivity [10]. LTE femtocells have been deployed in many markets. Driven by the increasing demand for mobile data, customers expect sufficient 5G indoor coverage in addition to 5G outdoor coverage. 5G femtocells can improve indoor coverage, offload traffic from macro gNBs, and allow for customized access control. In Release 19, 3GPP will study the overall RAN architecture, required functionalities and procedures for the support of 5G femtocell deployments. 3GPP will also investigate access control mechanism as well as access to local services from 5G femtocells.

A WAB node can be understood as a relay that holds full gNB function (including both CU and distributed unit (DU)) and mobile termination (MT) function, where the gNB function is used to communicate with UEs for access service and the MT function is used to communicate with another gNB for backhauling purpose. A WAB node is not the same as an IAB node as the latter only holds partial gNB function (i.e., gNB-DU) and MT function. The WAB node can act as a vehicle-mounted relay that provides access for UEs onboard aircrafts, cruise ships, helicopters, and vehicles in remote areas. In Release 19, 3GPP will study the support of WAB in the areas of architecture and protocol stack, WAB mobility within an existing RAN, inter-gNB and gNB to core network signaling, and resource multiplexing signaling enhancements.

A UE can also act as a relay known as sidelink relay. A UE-to-network sidelink relay connects a remote UE to the network, while a UE-to-UE sidelink relay connects one UE to another UE. In Release 18, 3GPP specified solutions for single-hop UE-to-network sidelink relay and single-hop UE-to-UE sidelink relay. However, single-hop sidelink relay may not be sufficient in terms of coverage extension for mission critical communications used by public safety personnel in out-of-coverage scenarios. Therefore, 3GPP RAN agreed that Release 19 will introduce the support of multi-hop sidelink relay with a focus on L2-based UE-to-network sidelink relay.

In addition to terrestrial topological evolutions, 3GPP has also been working on non-terrestrial topological evolutions, including NTNs that use satellites or high-altitude platforms to provide connectivity services [11]. After studies in Releases 15 and 16, 3GPP specified a set of basic features to enable NR operation over NTN in Release 17 and introduced enhancements for NR NTN in Release 18. This endeavor will continue in Release 19. 3GPP will define additional reference satellite payload parameters that will lead to a lower equivalent isotropic radiated power (EIRP) density per satellite beam than the nominal value considered in previous releases. Accordingly, downlink coverage enhancements will be investigated to cater for the reduced EIRP density per satellite beam. Considering that a large number of UEs will be in the coverage of a satellite, 3GPP will study how to increase uplink capacity by applying orthogonal cover codes to the discrete Fourier transform spread orthogonal frequency division multiplexing (DFT-s-OFDM) based physical uplink shared channel. As MBS feature is valuable for NR NTN, 3GPP will enhance MBS features for use in NR NTN by specifying the signaling of the intended service area via NR NTN. Another enhancement area is the support of regenerative payload, i.e., 5G system functions onboard the NTN platform. Compared to transparent payload supported in Releases 17 and 18, support of regenerative payload will make NTN deployments more flexible. Last but not least, NR NTN will be evolved to support RedCap UEs. 3GPP introduced support for RedCap UEs in Releases 17 and 18 to serve use cases such as industrial sensors, video surveillance, and wearables with terrestrial networks. The global coverage provided by NR NTN will benefit RedCap UEs.

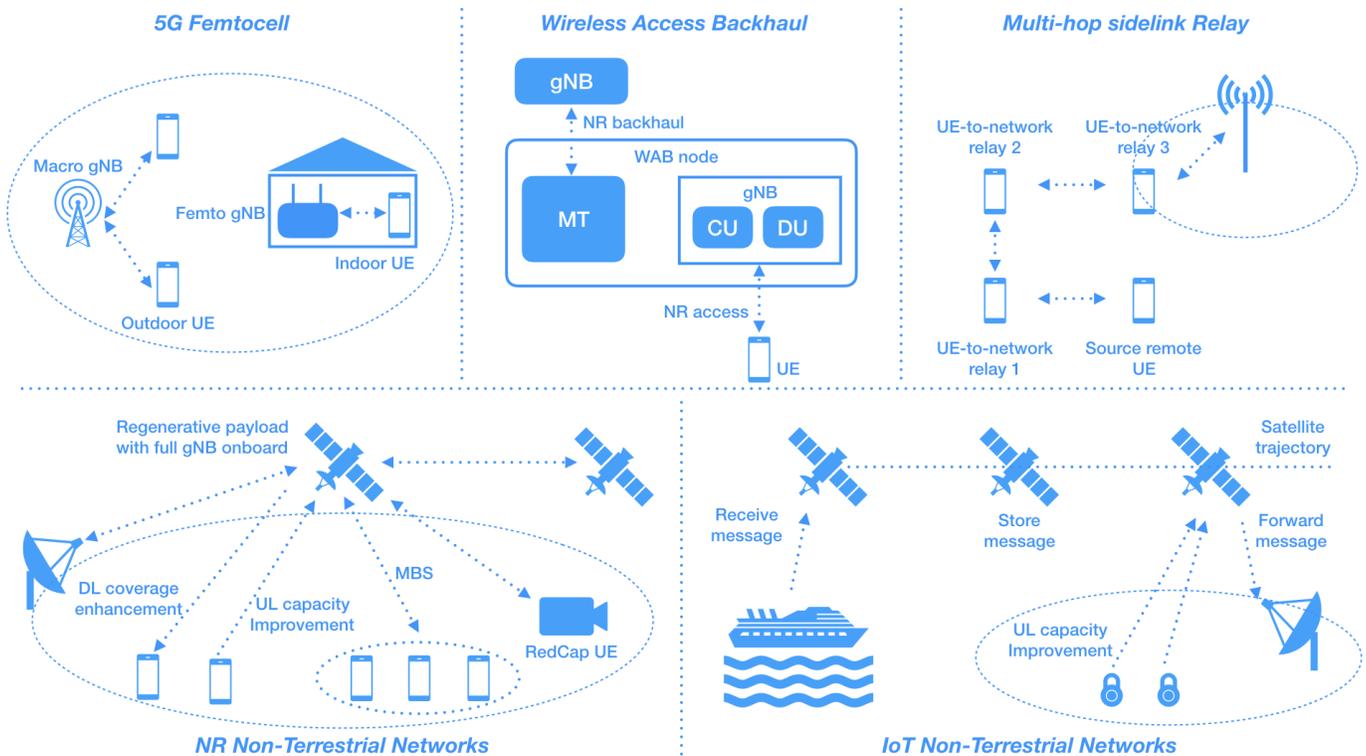

**Figure 3: An illustration of the key 3GPP Release-19 features related to further evolved 5G topology.**

It is worth noticing that there has also been interest in NTN-based massive internet of things (IoT) use cases using narrowband IoT (NB-IoT) and LTE machine type communication (LTE-M), known as IoT NTN. Similar to NR NTN, IoT NTN has been specified in Release 17 and enhanced in Release 18. 3GPP will introduce the support of store-and-forward satellite operation with regenerative payload for IoT NTN in Release 19. The store-and-forward satellite operation can offer delay-tolerant services in areas visited by the satellites but without ground infrastructure, such as mid-sea. 3GPP will also study how to increase uplink capacity through multiplexing of UEs by using orthogonal cover codes, as well as early data transmission enhancements.

## IV. ENERGY EFFICIENT 5G

Energy efficiency has been a key design target since the inception of 5G. For example, NR minimizes the always-on signals such as cell-specific reference signals that are always present in LTE. Over the releases, 3GPP has worked to further improve 5G energy efficiency and reduce power consumption for both gNB and UE. 3GPP Release 19 will continue working on energy efficiency related items including network energy savings, LP-WUS/LP-WUR, and ambient IoT, as illustrated in Figure 4.

Mobile network operators are concerned about 5G network energy consumption, despite that 5G offers a significant energy efficiency improvement (in terms of energy consumed per traffic unit) over previous generations of mobile networks [12]. Dense deployments, massive MIMO, large bandwidths, and many frequency bands have led to a high network energy consumption level in 5G, resulting in increased operating expense and carbon footprint. 3GPP conducted a study on network energy savings for NR in Release 18. In this study, 3GPP developed a base station energy consumption model and investigated a comprehensive set of network energy saving techniques in time, frequency, spatial, and power domains. After the study, 3GPP conducted normative work to specify several network energy saving enhancements in Release 18, including techniques for efficient adaptation of spatial elements such as antenna ports and active transceiver chains, cell discontinuous transmission (DTX) and discontinuous reception (DRX) mechanisms, among others. However, due to limited time and capacity in Release 18, 3GPP only specified a limited set of enhancements for network energy savings. Therefore, 3GPP continues conducting normative work in Release 19 to introduce more network energy saving techniques, including on-demand SSB in secondary cell for connected-mode UEs configured with CA, on-demand system information block type 1 (SIB1) for UEs in idle/inactive mode, and adaptation of common signal/channel transmissions.

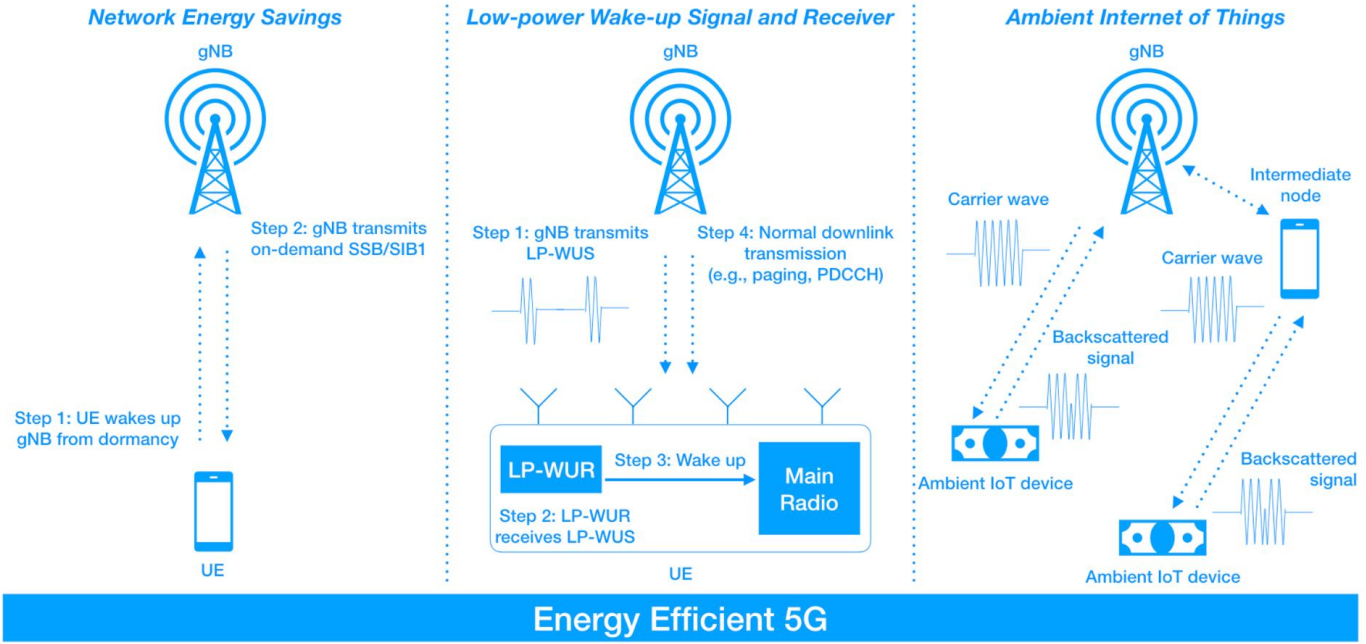

**Figure 4: An illustration of the key 3GPP Release-19 features related to energy efficient 5G.**

Besides network energy efficiency, 3GPP has also paid close attention to UE energy efficiency by specifying various UE power saving techniques over several releases. UE power saving is critical not only for smartphones to prolong battery life and improve user experience, but also for vertical applications where devices do not have rechargeable batteries. One vital UE power saving mechanism is DRX, whose basic idea is that UE periodically wakes up to monitor downlink signals in DRX ON durations and enters sleep mode for the rest of the time to save power. However, the periodic wake-up dominates UE power consumption in periods when there is no downlink signal for the UE. The power consumption can be reduced if the UE wakes up only when it receives a LP-WUS from the gNB. To achieve this, the UE can be equipped with a LP-WUR besides its main radio used for normal NR transmissions and receptions. The UE's main radio is turned on when the UE's LP-WUR detects a LP-WUS from the gNB. In Release 18, 3GPP investigated LP-WUR architectures, LP-WUS designs, and the required changes in L1 procedures and higher layer protocols. 3GPP also evaluated the impact of LP-WUS and LP-WUR on UE power saving, latency, coverage, network power consumption, and system overhead. In Release 19, 3GPP will specify a LP-WUS design. For UEs in idle/inactive mode, procedure and configuration that enable the LP-WUS to trigger the UE's main radio to monitor paging will be introduced; serving and neighbor cell measurements by the UE's main radio will be relaxed for RRM; and offloading serving cell RRM measurements from the UE's main radio to the LP-WUR, as well as low-power synchronization signal, will be specified. For UEs in connected mode, procedures that enable the LP-WUS to trigger the UE's main radio to monitor physical downlink control channel (PDCCH) will be introduced.

To support low-end IoT devices with ultra-low-power consumption (below 1 mW) and ultra-low complexity, 3GPP conducted a study in Release 18 to investigate ambient IoT use cases, deployment scenarios, and design targets [13]. Ambient IoT devices are either batteryless without energy storage capability or equipped with very limited energy storage that does not require manual replacement or recharge (e.g., energy harvesting of radio waves). One example use case is asset identification which primarily utilizes barcode and radio frequency identification (RFID) techniques today. The reading range of these two techniques is however limited to a few meters. The existing 3GPP low-power wide-area technologies such as NB-IoT and LTE-M cannot serve this IoT segment well because the complexity and power consumption of the ambient IoT devices will need to be orders-of-magnitude lower. The preliminary analysis conducted in Release 18 shows that it is feasible and beneficial to develop a new technology to support ambient IoT devices. In Release 19, 3GPP is committed to carrying out a further assessment of ambient IoT, including defining evaluation assumptions and studying solutions. In particular, the waveform characteristics of a carrier wave that causes an ambient IoT device's uplink transmission to be backscattered will be investigated among other physical layer aspects. Also, compact protocol stack, lightweight signaling procedure, RAN architecture aspects, and the interface between RAN and core network will be studied.

## V. AI-POWERED 5G

AI/ML has emerged as a powerful technique that can help to manage complex 5G networks efficiently and effectively. 3GPP has been incorporating AI/ML into 5G evolutions across multiple domains including core network, RAN, UE, and operations and management [14]. 3GPP Release 19 will continue to embrace AI/ML to make 5G-Advanced more intelligent, as illustrated in Figure 5.

Air interface is the core of any wireless communication system. In Release 18, 3GPP conducted a study on AI/ML for NR air interface, which was the first of its kind in the 3GPP history of mobile standards development. The study investigated a general framework for AI/ML as well as selected use cases including CSI feedback, beam management, and positioning. In Release 19, 3GPP will introduce support for the AI/ML general framework for one-sided AI/ML models by specifying signaling and protocol aspects related to life cycle management (LCM), where a one-sided model can be either UE-sided or network-sided. Evaluation results have shown that it is beneficial to adopt AI/ML-based beam management and AI/ML-based positioning. In Release 19, 3GPP will specify necessary signaling and mechanisms to enable AI/ML-based beam management and AI/ML-based positioning for both UE-sided model and network-sided model.

In addition to the aforementioned normative work, Release-19 work item on AI/ML for NR air interface also includes several study objectives to address the outstanding issues identified during the Release-18 study. One such area is CSI feedback, including spatial-frequency domain CSI compression and time domain CSI prediction at UE. CSI compression needs to use a two-sided AI/ML model with a UE-sided encoder and a network-sided decoder, leading to challenges in inter-vendor training collaboration. Furthermore, the performance gain demonstrated under Release-18 evaluation setup is not sufficient to justify the complexity/overhead of using a two-sided model for CSI compression. As a result, 3GPP will continue to study CSI compression in Release 19 to overcome the challenges of inter-vendor training collaboration and improve the tradeoff between performance and complexity/overhead. For CSI prediction studied in Release 18, there is a lack of results on the performance gain over non-AI/ML based approach as well as concerns on complexity. In Release 19, 3GPP will continue to study CSI prediction aspects including performance improvement and complexity issues, among others. Other study objectives include model identification, training data collection for UE-sided model, model transfer/delivery, and testability and interoperability issues.

At the NG-RAN architecture level, 3GPP conducted a work item in Release 18 on AI/ML-based network energy saving, load balancing, and mobility optimization. The work specified data collection enhancements and signaling support for these use cases. In Release 19, 3GPP will further enhance the support of these use cases using AI/ML. Candidate topics include split architecture support, mobility optimization for NR dual connectivity (DC), multi-hop UE trajectory across gNBs, energy cost prediction, and continuous minimization of drive tests (MDT) collection from the same UE across RRC states. In addition, 3GPP will study AI/ML-based network slicing, e.g., using AI/ML to optimize the resource allocation for network slicing. 3GPP will also study coverage and capacity optimization, whose objective is to utilize AI/ML to dynamically adapt cell/beam coverage (i.e., cell shaping).

AI/ML-based mobility optimization investigated at the NG-RAN architecture level in Release 18 was based on information available at network side. In Release 19, 3GPP will conduct a dedicated study on AI/ML for mobility in NR air interface that will further consider information available at UE side. Specifically, the study will investigate AI/ML-based RRM measurement prediction for both UE-sided model and network-sided model, as well as prediction of events (e.g., handover failure, radio link failure (RLF), and measurement events) for UE-sided model. The study will also examine potential specification impacts and the necessity of UE assistance information for network-sided model and. The evaluation of the AI/ML-based mobility will consider tradeoffs between handover performance and complexity. Additional evaluation aspects include testability, interoperability, and RRM requirements and performance.

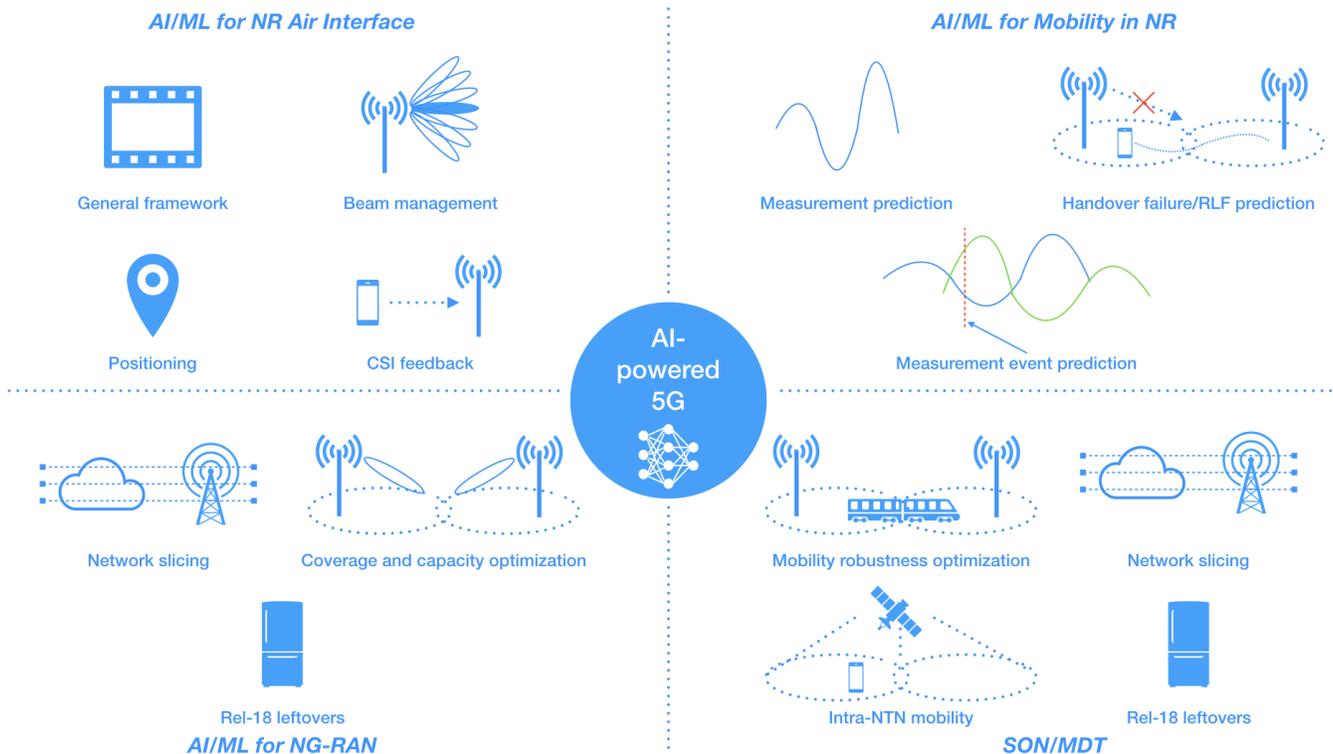

**Figure 5: An illustration of the key 3GPP Release-19 features related to AI-powered 5G.**

3GPP Release 19 will also enhance data collection for self-organizing network (SON)/MDT in NR standalone and multi-radio DC. SON enables network self-configuration and self-optimization, while MDT enables measurement data collection from normal UEs rather than drive tests. The Release-19 work will address SON/MDT features leftover from Release 18 and introduce support of SON/MDT enhancements for intra-NTN mobility and network slicing. In addition, 3GPP will look into mobility robustness optimization for several mobility mechanisms (e.g., LTM) by specifying inter-node information exchange and identifying necessary UE reporting.

## VI. TOWARD 6G

Though the primary focus of Release 19 is on the commercial deployment needs of 5G-Advanced, 3GPP is also starting to plan 6G work as the development of a new generation of mobile standards takes several years. At the December 2023 plenary meeting, 3GPP had a first discussion on 6G timeline. The expectation is that the discussion on 6G use cases and requirements will commence within 3GPP in 2024, followed by 6G technology studies in Release 20 from 2025 onwards. The following Release 21 will produce the first set of 3GPP 6G technical specifications.

Release 19 will serve as a bridge from 5G-Advanced to 6G. The key technology components in Release 19, such as AI/ML, massive MIMO enhancements, NTN integration with terrestrial networks, and network energy saving enhancements, will become precursors to several 6G building blocks [15]. In addition, 3GPP has planned two studies in Release 19 focusing on channel modeling for integrated sensing and communication (ISAC) and for 7-24 GHz spectrum to prepare for 6G standardization. These studies on channel modeling are crucial as the upcoming 6G studies and evaluations will require accurate channel models.

As a key 6G usage scenario, ISAC integrates sensing capabilities to obtain spatial information about connected or unconnected objects and their environments. Proper channel models are vital for evaluating ISAC use cases, such as object detection and tracking and distinguishment from non-targeted objects. There are, however, gaps in the existing 3GPP channel models to enable evaluation of sensing techniques. In Release 19, 3GPP will identify deployment scenarios for ISAC and define channel modelling for sensing using the existing 3GPP channel models as a starting point. The channel modeling for ISAC will consider modelling of sensing targets and background environment, including aspects such as radar cross-section, mobility, clutter patterns, and spatial consistency.

The 7-24 GHz spectrum is considered to be the main spectrum range for 6G. This frequency range has the potential to provide wide-area coverage with large bandwidths. Though the existing 3GPP channel models developed for 5G support channel modelling from 0.5 GHz to 100 GHz, limited channel measurements were available (especially for the 7-24 GHz spectrum) when those channel models were developed in 3GPP. Furthermore, larger antenna arrays are expected to be used in the 7-24 GHz spectrum compared to the sub-7 GHz spectrum, calling for additional considerations (e.g., near-field effects and

| Category | Approved/endorsed item | Study/work item/way forward | Responsible groups |
|---|---|---|---|
| **High Performing 5G** | RP-234007: NR MIMO Phase 5 | Work item | RAN 1, 2, 4 |
| | RP-234036: NR mobility enhancements Phase 4 | Work item | RAN 2, 1, 3, 4 |
| | RP-234035: Evolution of NR duplex operation: Sub-band full duplex (SBFD) | Work item | RAN 1, 2, 3, 4 |
| | RP-234080: XR (eXtended Reality) for NR Phase 3 | Work item | RAN 2, 1, 4 |
| **Further evolved 5G topology** | RP-234041: Study on additional topological enhancements for NR | Study item* | RAN 3, 2 |
| | RP-233998: Way forward on SL multi-hop relay | Way forward** | RAN 2 |
| | RP-234078: Non-terrestrial networks (NTN) for NR Phase 3 | Work item | RAN 2, 1, 3, 4 |
| | RP-234077: Non-terrestrial networks (NTN) for Internet of Things (IoT) Phase 3 | Work item | RAN 2, 1, 3 |
| **Energy efficient 5G** | RP-234065: Enhancements of network energy savings for NR | Work item | RAN 1, 2, 3, 4 |
| | RP-234056: Low-power wake-up signal and receiver for NR | Work item | RAN 1, 2, 3, 4 |
| | RP-234058: Study on solutions for ambient IoT (Internet of Things) in NR | Study item*** | RAN 1, 2, 3, 4 |
| **AI-powered 5G** | RP-234039: Artificial intelligence (AI)/machine learning (ML) for NR air interface | Work item | RAN 1, 2, 3, 4 |
| | RP-234054: Study on enhancements for artificial intelligence (AI)/machine learning (ML) for NG-RAN | Study item* | RAN 3 |
| | RP-234055: Study on artificial intelligence (AI)/machine learning (ML) for mobility in NR | Study item | RAN 2, 4 |
| | RP-234038: Data collection for SON (self-organising networks)/MDT (minimization of drive tests) in NR standalone and MR-DC (multi-radio dual connectivity) Phase 4 | Work item | RAN 3, 2 |
| **Toward 6G** | RP-234069: Study on channel modelling for integrated sensing and communication (ISAC) for NR | Study item | RAN 1 |
| | RP-234018: Study on channel modelling enhancements for 7-24 GHz for NR | Study item | RAN 1 |

*Note 1: This study item is expected to be followed by a work item in the Release-19 timeframe.
**Note 2: The work item description on multi-hop sidelink relay is expected to be approved at the September 2024 RAN plenary.
***Note 3: This study item targets 12-month completion and will be followed by a work item or a continuation of the study in Release 19.

**Table 1: A summary of 3GPP RAN Release-19 package approved at the December 2023 RAN plenary meeting.**

spatial non-stationarity) in the channel modeling. In Release 19, at least for the 7-24 GHz spectrum, 3GPP will validate its existing channel models with measurement data and, if necessary, adapt the channel models to take into account near-field propagation and spatial non-stationarity.

## VII. CONCLUSIONS

5G-Advanced represents the next significant phase in the development of 5G technology. As outlined in this article, Release 19–the second release of 5G-Advanced–will primarily focus on the commercial deployment needs by further enhancing performance, evolving network topology, improving energy efficiency, and utilizing AI/ML. Table 1 provides a summary of the 3GPP Release-19 package discussed in this article.

The emerging use cases and requirements will introduce challenges that go beyond the capabilities of 5G-Advanced, calling for 6G. To this end, Release 19 will act as a stepping stone toward 6G by initiating channel modeling studies for ISAC and 7-24 GHz spectrum. The innovative work carried out for 5G-Advanced in Release 19 will provide the baseline for 6G standardization.